\documentclass[12pt]{article}
\usepackage{graphicx}

\title{\bf Theory of Mixing and CP violation}

\author{Alexander Lenz
        \\
        IPPP, Durham University, UK\\
        E-mail: alexander.lenz@durham.ac.uk}

\begin{document}

\clearpage\maketitle 
\thispagestyle{empty}

\begin{centering}
\subsection*{Abstract}
\end{centering}
%\begin{sciabstract}
We review the current status of B-mixing observables and point out the
  crucial importance of a control of the hadronic uncertainties for ruling out or confirming  hints of
  BSM physics. In addition we introduce a rating system for theory predictions
  for lifetimes and mixing observables, that classifies the quality of the corresponding
  SM values ranging from no star to ****.
% \end{sciabstract}

\newpage

\section{Introduction}
\begin{figure*}
  \begin{center}
\includegraphics[width=0.90 \textwidth]{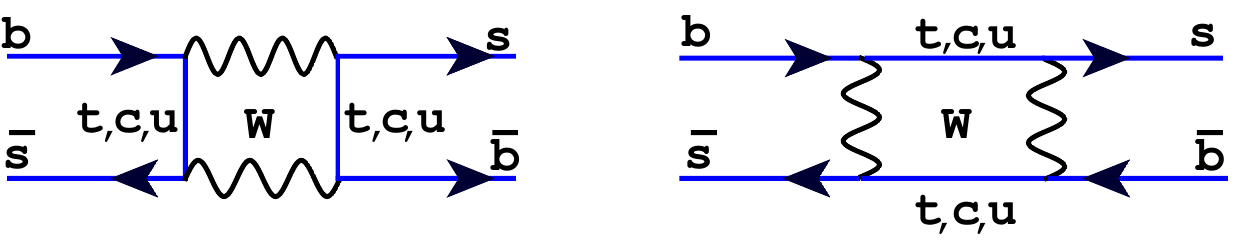}
\caption{\label{box} Standard Model  diagrams for the transition between $B_s$ and $\bar{B}_s$ mesons.}
\end{center}
  \end{figure*}

In the Standard Model (SM) mixing of neutral $B_q$-mesons is governed by the famous box-diagrams,
with internal $W$-bosons
and internal $up$-, $charm$- and $top$-quarks, see Fig. \ref{box} for the case of $B_s$-mesons -
for a more detailed introduction
into $B$-mixing, see e.g. \cite{Artuso:2015swg}.
The contribution of internal on-shell particles
(only the $charm$- and the $up$-quark can contribute) is denoted by
$\Gamma_{12}^q$; the contribution of internal off-shell particles
(all depicted particles can contribute) is denoted by $M_{12}^q$. In the $B$-system there are simple
relations\footnote{This holds not for $D$-mixing, see e.g.
\cite{Jubb:2016mvq,Bobrowski:2010xg,Nierste:2009wg}.}
between $\Gamma_{12}^q$, $M_{12}^q$ and the physical observables mass difference $\Delta M_q$,
the decay rate difference $\Delta \Gamma_q$ and the semi-leptonic asymmetries $a_{sl}^q$:
\begin{equation}
  \Delta M_q  \approx  2 \left|M_{12}^q \right| \, ,
  \hspace{1cm}
  \Delta \Gamma_q \approx  2 \left|\Gamma_{12}^q \right| \cos \phi_{12}^q \, ,
 \hspace{1cm}
  a_{sl}^q  \approx  \left| \frac{\Gamma_{12}^q }{M_{12}^q} \right| \sin \phi_{12}^q \, .
\end{equation}
The  calculation of $M_{12}^q$ gives
\begin{eqnarray}
M_{12}^q & = & \frac{G_F^2}{12 \pi^2} \lambda_t^2 M_W^2 S_0(x_t) B f_{B_q}^2  M_{B_q} \hat{\eta }_B\, ,
\label{M12}
\end{eqnarray}
where $\lambda_t$ denotes the CKM elements $V_{tq}^* V_{tb}$ and the
Inami-Lim function  $S_0$ \cite{Inami:1980fz} contains the result of the 1-loop box diagram in the SM.
The bag parameter $B$ and the decay constant $f_{B_q}$
quantify the hadronic contribution to $B$-mixing, the uncertainties of their numerical values make
up the by far biggest uncertainty in the SM prediction of the mass difference. Perturbative
2-loop QCD corrections have been calculated by
\cite{Buras:1990fn} and they are compressed in the factor $\hat{\eta }_B $.
The calculation of $\Gamma_{12}^q$ is more involved and is based on the Heavy Quark Expansion (HQE)
(see \cite{Lenz:2014jha} for a review and the original references).
According to the HQE the total decay rate of a
heavy hadron can be expanded in the inverse of the heavy quark mass as
\begin{eqnarray}
  \frac{1}{\tau} = \Gamma & = & \Gamma_0 + \frac{\Lambda^2}{m_b^2}\Gamma_2
  + \frac{\Lambda^3}{m_b^3}\Gamma_3
  + \frac{\Lambda^4}{m_b^4}\Gamma_4 + ... \, .
\end{eqnarray}
The hadronic scale $\Lambda$ is of order $\Lambda^{QCD}$, its numerical value has to be determined
by direct computation.
For hadron lifetimes it turns out that the dominant correction to $\Gamma_0$ is the third term
$\Gamma_3$. Each of the $\Gamma_i$'s can be split up in a perturbative part and non-perturbative
matrix elements - it can be
formally written as
\begin{eqnarray}
  \Gamma_i & = & \left[ \Gamma_i^{(0)} + \frac{\alpha_S}{4 \pi} \Gamma_i^{(1)}
                                      + \frac{\alpha_S^2}{(4 \pi)^2} \Gamma_i^{(2)} + ... \, ,
    \right] \langle O^{d=i+3} \rangle
\end{eqnarray}
where $\Gamma_i^{(0)}$ denotes the perturbative LO-contribution, $\Gamma_i^{(1)}$ the NLO one and so on;
$\langle O^{d=i+3} \rangle$ is the non-perturbative matrix element of $\Delta B = 0$ operators
of dimension $i+3$. The mixing quantity $\Gamma_{12}^q$ obeys a very similar HQE, but now the operators change the $b$-quantum number by two units, $\Delta B = 2$:
\begin{eqnarray}
 \Gamma_{12} & = & 
  \frac{\Lambda^3}{m_b^3}\Gamma_3
  + \frac{\Lambda^4}{m_b^4}\Gamma_4 + ... \, .
\end{eqnarray}
\section{Current Status}
\begin{figure*}
  \begin{center}
\includegraphics[width=0.90 \textwidth]{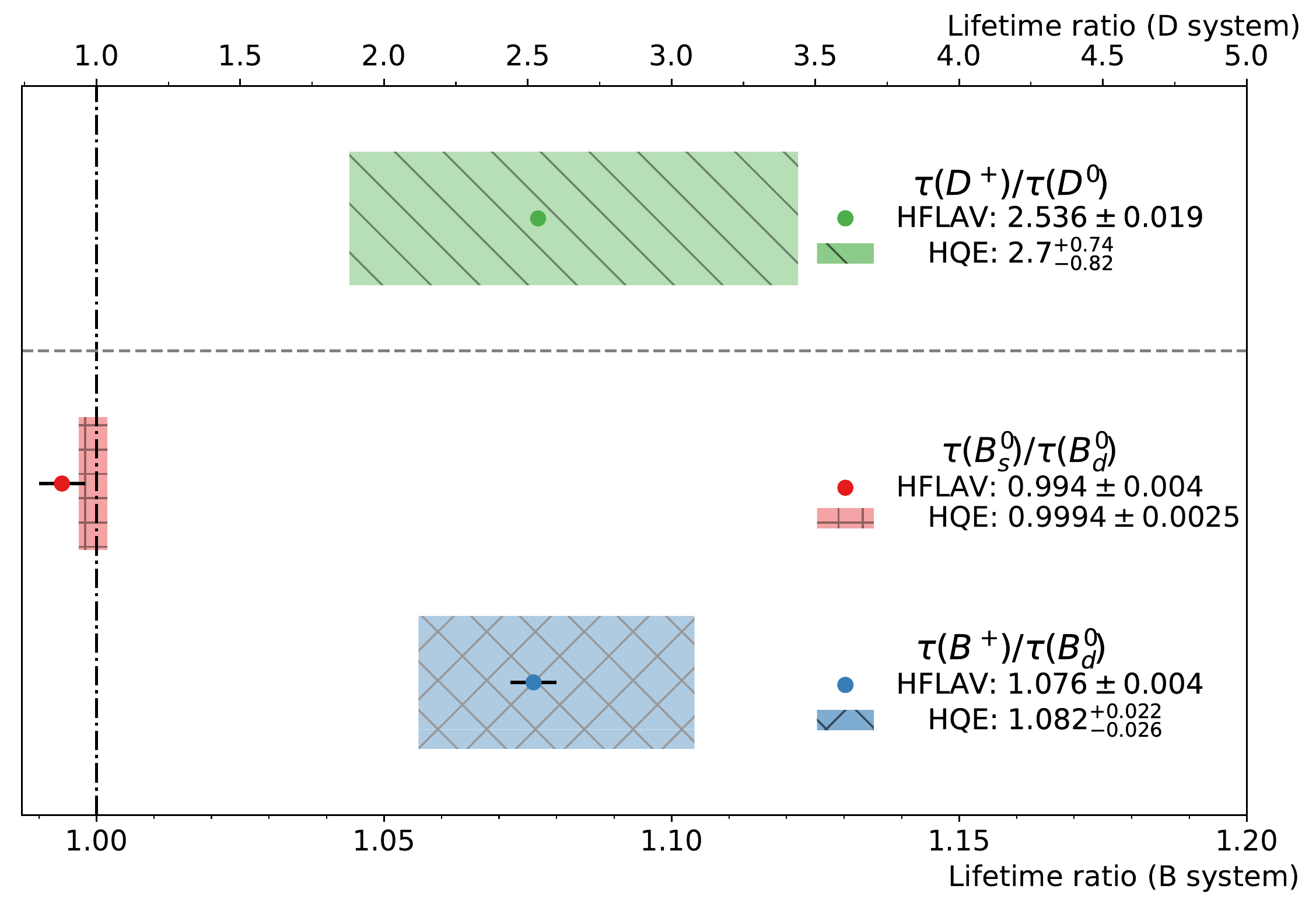}
\caption{\label{lifetime} Comparison of the most solid SM predictions for heavy lifetimes with experiment.}
  \end{center}
\end{figure*}

We introduce in this section a rating system for the robustness of lifetime and mixing
predictions. Any calculation of a perturbative term ($\Gamma_i^{(j)}$) or a non-perturbative matrix element ($\langle O^{d=k} \rangle$)
gets a $''+''$; if the calculation is confirmed by an independent collaboration it gets a $''++''$. In the case of non-perturbative matrix
elements one can even gain a $''+++''$ for two independent lattice evaluations and one sum rule evaluation.
A missing non-perturbative matrix element of dimension 6 is punished by a $''- -''$ contribution.
Non-perturbative estimates different from lattice or sum rules (like quark models) will be
valued by a $''0''$. Partial perturbative calculations will be rated with a $''+/2''$.
The possible number of 15 ``+'' will be classified in 5 categories:
**** (at least 12 ``+''),
***  (at least  8 ``+''),
**   (at least  4 ``+''),
*    (at least  2 ``+'') and
no star for 1 or less ``+''.
\\
For the lifetimes of heavy hadrons we get the following overview:
\begin{displaymath}
  \begin{array}{|c||c|c|c|c||c|c|c||ll|}
    \hline
    Obs. & \Gamma_3^{(0)} & \Gamma_3^{(1)} & \Gamma_3^{(2)} & \langle O^{d=6} \rangle & \Gamma_4^{(0)} & \Gamma_4^{(1)} & \langle O^{d=7} \rangle & \sum & 
    \\
    \hline
    \hline
    \tau (B^+) / \tau (B_d)         & ++ & ++           & 0  & +      & ++  & 0 & 0 & ** &(7+)
    \\
    \hline
    \tau (B_s) / \tau (B_d)        & ++ & ++            & 0  & \frac{+}{2}      & ++ & 0 & 0 & ** &(6.5+) 
    \\
     \hline
     \tau (\Lambda_b) / \tau (B_d)  & ++ & \frac{+}{2}  & 0  &  \frac{+}{2}     & + & 0 & 0 & ** & (4+)
     \\
     \hline
     \tau (b-baryon) / \tau (B_d)  & ++ & 0             & 0  & 0      & + & 0 & 0 & * & (3+)
     \\
     \hline
     \tau (B_c)                     & + & 0             & 0  & +      & 0 & 0 & 0 &  *  & (2+)
     \\
     \hline
     \tau (D^+) / \tau (D^0)       & ++ & ++            & 0  & +      & ++ & 0 & 0 & ** &(7+)
    \\
     \hline
     \tau (D_s^+) / \tau (D^0)     & ++ & ++            & 0  & \frac{+}{2}        & ++  & 0 & 0 & ** &(6.5+)
    \\
    \hline
     \tau (c-baryon) / \tau (D^0)  & ++ & 0             & 0  & 0       & + & 0 & 0 &  * & (3+)
     \\
    \hline
 \end{array}
\end{displaymath}
The LO-QCD part $\Gamma_3^{(0)}$ was first done with the full charm quark mass dependence in 1996
by Uraltsev \cite{Uraltsev:1996ta}  and Neubert and Sachrajda \cite{Neubert:1996we}. For the $B_c$-meson
one has to estimate also the leading HQE term $\Gamma_0 $ - the full estimate of the lifetime was done  by Beneke and Buchalla \cite{Beneke:1996xe} - to
some extent this quantity does not perfectly fit in our list.
The NLO-QCD corrections $\Gamma_3^{(1)}$ to $B^+$, $B_d$ and $B_s$ were done by \cite{Beneke:2002rj}
and the Rome group \cite{Franco:2002fc} - the Rome group also presented part of the NLO-QCD corrections
for the $\Lambda_b$. In the charm system the NLO-QCD corrections were done by \cite{Lenz:2013aua} for $D$-mesons.
The dimension 6 matrix elements for mesons (except for small corrections arising in $B_s$ and $D_s$) were recently calculated
via HQET sum rules \cite{Kirk:2017juj} - here a complementary lattice evaluation would be very important, either for looking for BSM effects
in the very precisely predicted ratio $\tau (B_s)/ \tau(B_d)$
- this could point towards new effects in hadronic tree-level decays \cite{Jager:2017gal} - , or for testing
the convergence of the HQE in the $b$- and in particular in the $charm$-system. For baryons we do not have a complete first principle determination
of the non-perturbative matrix elements - there are sum rule determinations of the condensate contribution for the $\Lambda_b$
\cite{Colangelo:1996ta} -  we have, however, some estimates \cite{Lenz:2014jha,Cheng:2018rkz} of the size of the matrix elements using spectroscopy
as an input (based on \cite{Rosner:1996fy}).
LO dimension 7 contributions to $B^+$, $B_s$, $B_d$ and $\Lambda_b$ were done in \cite{Gabbiani:2003pq}. These authors also
considered dimension 8 contribution, but since there are operators arising where we even cannot use vacuum insertion approximation, we did not
include these corrections in our list. There are unpublished calculations of the dimension 7 terms to  $B^+$, $B_s$ and $B_d$ by Uli Nierste
and myself, that agree with \cite{Gabbiani:2003pq}, therefore the ``++'' in the table. Perturbative dimension 7 contributions to D mesons
were determined in \cite{Lenz:2013aua} and to charmed baryons in \cite{Cheng:2018rkz}. So far there exists no non-perturbative determination
of the matrix elements of dimension 7 operators.
In Fig. \ref{lifetime}, taken from \cite{Kirk:2017juj},  we compare the most solid SM predictions for heavy lifetimes with experiment and find an
excellent agreement.
\\
The SM prediction for the mass difference is completely dominated by the non-perturbative input for the matrix element of the
dimension 6 operator with a V-A Dirac structure.
Depending on this input we get the range of predictions for the mass difference in the $B_s$-system as indicated in Table \ref{ME}, taken
from \cite{DiLuzio:2017fdq}.
\begin{center}
  \begin{table*}
    \begin{center}
\begin{tabular}{|c||c|c|}
\hline
$\mbox{Source}$     & $f_{B_s} \sqrt{\hat{B}} $           & $\Delta M_s^{\rm SM} $   
\\
\hline
\hline
HPQCD14 \cite{Dowdall:2014qka}    &  $ (247 \pm 12) \; {\rm MeV} $  & $(16.2 \pm 1.7) \, \mbox{ps}^{-1} $
\\
\hline
HQET-SR \cite{Kirk:2017juj}    &  $ (261 \pm 8) \; {\rm MeV} $  & $(18.1 \pm 1.1)\, \mbox{ps}^{-1} $
\\
\hline
ETMC13 \cite{Carrasco:2013zta}    &  $ (262 \pm 10) \; {\rm MeV} $  & $(18.3 \pm 1.5) \, \mbox{ps}^{-1} $
\\
\hline
HPQCD09 \cite{Gamiz:2009ku} = FLAG13 \cite{Aoki:2013ldr}  &  $ (266 \pm 18) \; {\rm MeV} $  & $(18.9 \pm 2.6)\, \mbox{ps}^{-1} $
\\
\hline
\textbf{FLAG17} \cite{Aoki:2016frl} & \textbf{ $(274 \pm 8) \; {\rm MeV}$} & $(20.01 \pm 1.25)\, \mbox{ps}^{-1}$
\\
\hline
Fermilab16 \cite{Bazavov:2016nty}  &  $ (274.6 \pm 4) \; {\rm MeV} $  & $(20.1 \pm 0.7) \, \mbox{ps}^{-1} $
\\
\hline
HPQCD06  \cite{Dalgic:2006gp} &  $ (281 \pm 20) \; {\rm MeV} $  & $(21.0 \pm 3.0) \, \mbox{ps}^{-1} $
\\
\hline
RBC/UKQCD14  \cite{Aoki:2014nga} &  $ (290 \pm 20)\; {\rm MeV} $  & $(22.4 \pm 3.4) \, \mbox{ps}^{-1} $
\\
\hline
Fermilab11 \cite{Bouchard:2011xj}  &  $ (291 \pm 18) \; {\rm MeV} $  & $(22.6 \pm 2.8) \, \mbox{ps}^{-1} $
\\
\hline
\end{tabular}
\end{center}
    \caption{List of predictions for the non-perturbative parameter $f_{B_s} \sqrt{\hat{B}}$ and the corresponding SM prediction
      for $\Delta M_s$. The current FLAG average is dominated by the FERMILAB/MILC value from 2016.}
\label{ME}
\end{table*}
\end{center}
For the SM predictions of the decay rate differences in the $B_d$ and $B_s$-system we get the following list:
\begin{displaymath}
  \begin{array}{|c||c|c|c|c||c|c|c||l|}
    \hline
    Obs. & \Gamma_3^{(0)} & \Gamma_3^{(1)} & \Gamma_3^{(2)} & \langle O^{d=6} \rangle & \Gamma_4^{(0)} & \Gamma_4^{(1)} & \langle O^{d=7} \rangle  & \sum 
    \\
    \hline
    \hline
    \Gamma_{12}^s& ++ & ++  & \frac{+}{2}& ++ & ++ & 0 & 0 & 8.5+ (***)
    \\
    \hline
    \Gamma_{12}^d & ++ & ++ & 0          & +++ & ++ & 0 & 0 & 9+ (***)
    \\
     \hline
  \end{array}
\end{displaymath}
The NLO-QCD corrections $\Gamma_3^{(1)}$ have been calculated in \cite{Beneke:1998sy,Beneke:2003az,Ciuchini:2003ww}, recently also a part of the NNLO-QCD
has been determined \cite{Asatrian:2017qaz}. At dimension 6 two additional operators
to the one appearing in the mass difference are arising. We have currently a HQET sum rule determination for $B_d$ mesons
\cite{Grozin:2016uqy,Kirk:2017juj}
and lattice determinations from 2016 \cite{Bazavov:2016nty} ($N_f=2+1$) and 2013 \cite{Carrasco:2013zta} ($N_f=2$). The dimension 7
perturbative part has been determined already in 1996 by Buchalla and Beneke \cite{Beneke:1996gn} for $B_s$ and in \cite{Dighe:2001gc} for $B_d$.
For numerical values of the mixing observables see e.g. the {\it aggressive scenario} of \cite{Jubb:2016mvq}
\begin{eqnarray}
  \Delta \Gamma_s = (0.098 \pm 0.014) \mbox{ps}^{-1} \, ,  && a_{sl}^s = (2.27 \pm 0.25) \cdot 10^{-5} \, ,
  \\
  \Delta \Gamma_d = (2.99 \pm 0.52) \cdot 10^{-3} \mbox{ps}^{-1} \, ,  && a_{sl}^d = -(4.90 \pm 0.54) \cdot 10^{-4} \, .
\end{eqnarray}

\section{One constraint to kill them all}
\begin{center}
\begin{figure*}
  \begin{center}
    \includegraphics[width=0.80 \textwidth]{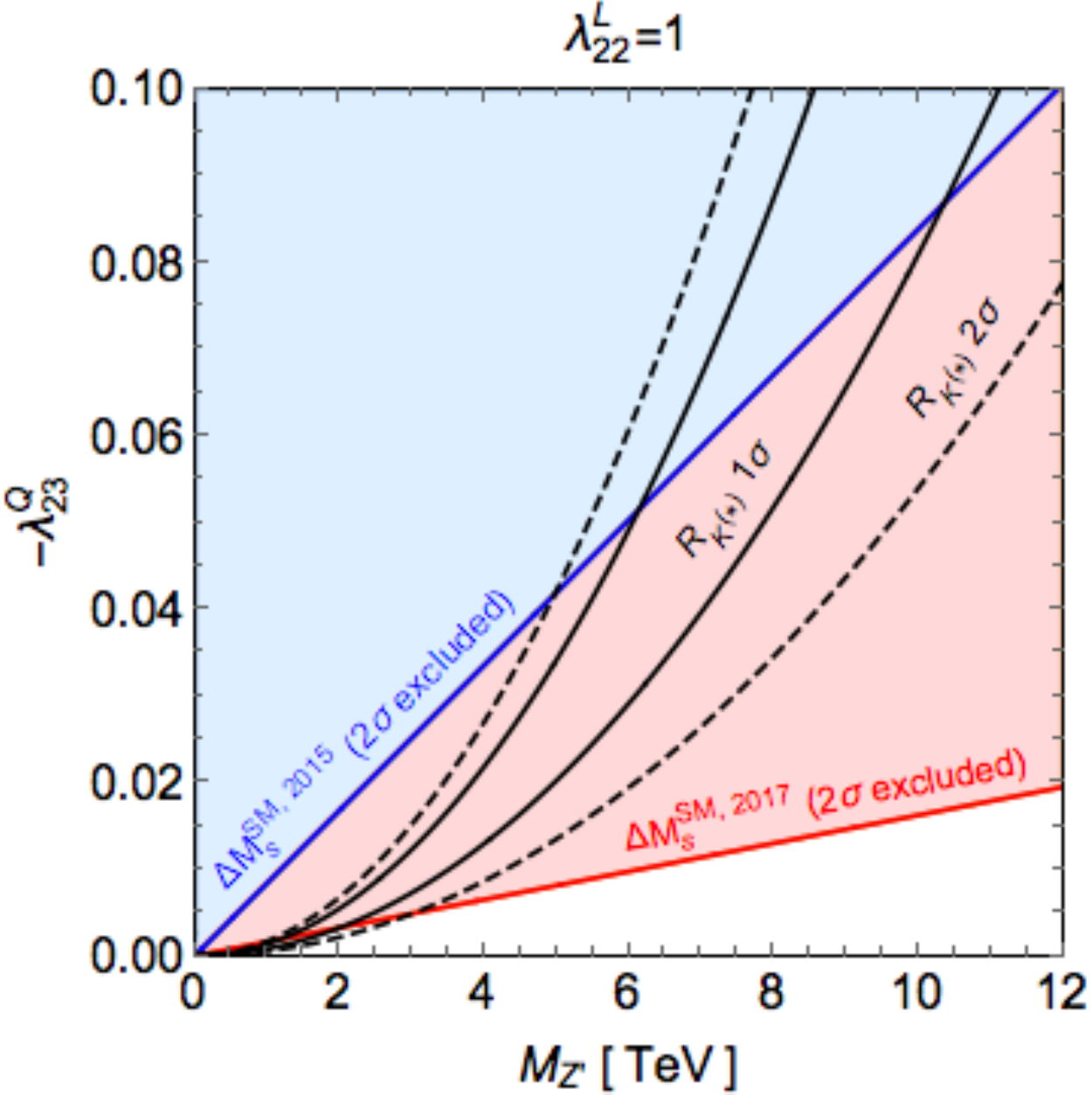}
\caption{\label{bound} Allowed parameter space of $Z'$ models that try to explain the B anomalies.}
  \end{center}
\end{figure*}
\end{center}

The importance of the precise value of SM predictions and a strict control of the corresponding uncertainties was highlighted recently
in \cite{DiLuzio:2017fdq}. Lepto-quarks and $Z'$ models are popular explanations of the B anomalies\footnote{Due to time and space restrictions I will not attempt
  to cite
  the numerous relevant papers in that field.}; these new models would also affect B-mixing - in the case of $Z'$ models  already at tree-level.
In Fig. \ref{bound} (from \cite{DiLuzio:2017fdq}) we show the allowed parameter range for a $Z'$ model: in order to explain e.g. $R_{K^{(*)}}$ the mass of the $Z'$
and the coupling to the $b$- and $s$-quark should lie within the black parabola-like shape (the 1 sigma bound is a solid line, the 2 sigma one a dotted line).
Taking the FLAG inputs from 2013 for the mass difference one can exclude the blue region. Taking the new FLAG average, that is dominated by the 2016 FNAL/MILC we are left with the red exclusion region and almost all of the possible parameter space of the $Z'$ model is excluded.

\section{Conclusion}
We presented an overview of the current theoretical status of lifetime and mixing predictions.
$\Delta \Gamma_q$ and $a_{sl}^q$ get the highest ranking (***). $\Gamma_{12}^s$ is slightly less precise known, because
 the HQET sum rule calculation does not include yet $m_s$-effects.
To improve further the reliability of these predictions one needs a non-perturbative determination
of the dimension 7 matrix elements (first steps have been done in \cite{Davies:2017jbi}) and perturbative evaluations of the $\alpha_s^2$- and
$\alpha_s/m_b$-corrections.
The next solid class of theoretical rigidness is (**) for $\tau (B^+) / \tau (B_d)$ and  $\tau (D^+) / \tau (D^0)$. Here an independent lattice determination
of the dimension 6 matrix elements is urgently needed.
$\tau (B_s) / \tau (B_d)$ and  $\tau (D_s^+) / \tau (D^0)$ is slightly less well known, because the $m_s$ corrections
to the HQET sum rule are not yet available. Finally $\Lambda_b$ is considerably less well-known but still a (**) - here we need urgently a first non-perturbative
determination of the dimension 6 matrix element.
Finally we have the (*) class, which one should consider more an estimate than a precise SM prediction with well-defined uncertainties.
We pointed out the crucial significance of a precise non-perturbative input for $\Delta M_q$ and related BSM studies - here an independent
$N_f = 2+1$ or $N_f = 2+1+1$ confirmation of the FNAL/MILC result of 2016 would be desirable.

{\bf \large Acknowledgement}
\\
I would like to thank the organisers and Luca Silvestrini for inviting me, Thomas Rauh for critical remarks on my classification scheme, Matthew Kirk,
Luca Di Luzio and Thomas Rauh for the pleasant collaboration and STFC for support via the IPPP research grant.

\end{document}